\documentclass[12pt prd,a4paper,dlspace,twocolumn
]{revtex4}
\usepackage{color}
\usepackage{epsfig}
\newcommand{\be}{\begin{equation}}
\newcommand{\ee}{\end{equation}}
\newcommand{\ba}{\begin{eqnarray}}
\newcommand{\ea}{\end{eqnarray}}
\newcommand{\nn}{\nonumber\\}
\begin{document}
\title{A geometric approach to correlations and quark number susceptibilities}
\author{Stefano Bellucci$^{a}$}
\email{bellucci@lnf.infn.it}
\author{Vinod Chandra$^{b}$}
\email{vinodc@theory.tifr.res.in}
\author{Bhupendra Nath Tiwari$^{a}$}
\email{tiwari@lnf.infn.it} \affiliation{$^{a}$ INFN-Laboratori
Nazionali di Frascati, Via E. Fermi 40, 00044 Frascati, Italy.}
\affiliation{$^{b}$ Department of Theoretical Physics, Tata
Institute of Fundamental Research, Homi Bhabha Road Mumbai-400005,
India.}
\begin{abstract}
We study the thermodynamic geometry arising from the free energy
for the 2- and 3-flavor finite temperature hot QCD near the
critical temperature. We develop a geometric notion for QCD
thermodynamics, relating it with the existing microscopic
quantities, e.g. quark-number susceptibility, which appears
naturally within an approximately self-consistent resummation of
perturbative QCD. We further incorporate thermal fluctuations in
the free energy, thus yielding the geometric properties of local
and global chemical correlations. These investigations are
perturbative in nature. Nevertheless, one could apply the same
line of thought for the geometric realization of underlying quark
susceptibilities, either in the fabric of lattice QCD or in that
of non-perturbative QCD.

\vspace{2mm} {\bf Keywords}: Hot QCD, Thermodynamic Geometry,
Quark-number susceptibility.

\end{abstract}
\maketitle
\section{Introduction}
Thermodynamic geometry \cite{wein,rup1,rup2} has emerged as an
important concept to understand the nature of correlations and
physics of phase transition. It has been studied in the context of
diverse thermodynamic systems: starting from the mixture of gases
to the thermodynamics of black holes in string theory
\cite{bnt,rup3,aman}.

Quantum chromodynamics (QCD) on the other hand is a well
celebrated physical theory of the strong interaction. QCD at high
temperature (hot QCD) plays a crucial role in understanding the
nature of the confinement-deconfinement phase transitions
(hadronic phase to deconfined phase (quark-gluon plasma phase).
The physics of strong interactions close to a phase transition
point, $T_c$,  is very complex and rich in this context.

Here, we have analyzed it in the important regime of the hot QCD
by setting up firstly the thermodynamic geometric notion of the
fluctuations of the hot QCD free energy, near the $T_c$ and then
investigated its connection with the existing studies of the
quark-number susceptibilities \cite{hep-ph/0110369v3,
hep-ph/0101103,hep-ph/0105183}. The free-energy of QCD has been
constructed employing the quasi-particle understanding of the 2-
and 3-flavor hot QCD  and the physical nature has been analyzed
near the $T_c$. We have further incorporated the role of thermal
fluctuations and thereby studied their impact from the notion of
the thermodynamic geometry.

The main aim here is to investigate the relation between the
covariant thermodynamic geometry and the quark-number
susceptibility tensor. Both of these have recently received
considerable attention in the black hole \cite{bnt,rup3,aman} and
QCD thermodynamics \cite{hep-ph/0110369v3,hep-ph/0101103,
hep-ph/0105183}.
\section{Thermodynamic Geometry}
Let us now present a brief review of thermodynamic geometry from
the perspective of the hot QCD. Let $ F(\mu_i, T) $ be the free
energy for given chemical potentials, and temperature $ \lbrace F,
\mu_i, T \rbrace$. Near the transition temperature, the free
energy of hot QCD depends on the quark chemical potentials
\cite{transition}. In this case, the $T_C$ as the function of the
quark chemical potentials, results in the following expression \be
\label{eq11} \frac{T_c(\mu)}{T^0_c}=1+\sum^{2}_{i=1} \tilde
{a}_i\frac{\mu_i^2}{(T^0_c)^2},\ee where, $ \tilde{ a}_1=\tilde
{a}_2=-0.07$ for the 2-flavor case and $\tilde{a}_1=
\tilde{a}_2=-0.114$ for the 3-flavor QCD. In general, considering
the high temperature QCD as the effective quasi-particle system,
we have constructed the thermodynamic geometric perspective of hot
QCD \cite{ourpaper} with the following free energy \ba
\label{eq12} F(\mu_1, \mu_2):&=& a_0(b+ a_1 \mu_1^2+ a_2
\mu_2^2)^4+ (b+ a_1 \mu_1^2+ a_2 \mu_2^2)^2\nn && \times (c_1
\mu_1^2+ c_2 \mu_2^2) + c_4^{(1)} \mu_1^4+ c_4^{(2)} \mu_2^4, \ea
where $b=T^0_c$, $a_1=\tilde{a_1}/b$ and $a_2=\tilde{a_2}/b$.
Here, the dimension of $a_1, a_2$ is in the unit of $b^{-1}$(GeV).
The free energy is measured in units of $GeV^{4}$.

The free energy with the incorporation of the thermal fluctuations
rectifies to the following expression \ba \label{eq14} F(\mu_1,
\mu_2):&=& a_0+ \frac{c_1 \mu_1^2+ c_2 \mu_2^2}{(b+ a_1 \mu_1^2+
a_2 \mu_2^2)^2}\nn &&+\frac{1}{2}\ln(\frac{c_1 \mu_1^2+ c_2
\mu_2^2}{(b+ a_1 \mu_1^2+ a_2 \mu_2^2)^2}). \ea

Following \cite{ourpaper}, the chemical potentials $\mu_i$ are
taken in the units of $GeV$. Thus, the components of the Weinhold
metric tensor \cite{wein} are in the units of $GeV^2$ and the
thermodynamic curvature is in $GeV^4$. In this study, the
components of the metric tensor are, \ba \label{eq2} g_{\mu_1
\mu_1} =\frac{\partial^2 F}{\partial \mu_1^2};\ \ g_{\mu_1 \mu_2}
=\frac{\partial^2 F}{{\partial \mu_1}{\partial \mu_2}};\ \
g_{\mu_2 \mu_2} =\frac{\partial^2 F}{\partial \mu_2^2}. \ea

Subsequently, the determinant of the metric tensor and scalar
curvature \cite{ourpaper} are given by, \ba \label{eq3} \Vert g
\Vert&=& F_{\mu_1\mu_1}F_{\mu_2 \mu_2}- F_{\mu_1 \mu_2}^2, \nn
R&=& -\frac{1}{2 \Vert g \Vert^2} \bigg(F_{\mu_2
\mu_2}F_{\mu_1\mu_1\mu_1} F_{\mu_1 \mu_2 \mu_2} -
F_{\mu_2\mu_2}F_{\mu_1\mu_1\mu_2}^2 \nn &&+ F_{\mu_1 \mu_2}
F_{\mu_1\mu_1\mu_2} F_{\mu_1\mu_2\mu_2}+
F_{\mu_1\mu_1}F_{\mu_1\mu_1\mu_2} F_{\mu_2\mu_2\mu_2}\nn &&-
F_{\mu_1\mu_2}F_{\mu_1\mu_1\mu_1} F_{\mu_2\mu_2\mu_2}-
F_{\mu_1\mu_1}F_{\mu_1\mu_2\mu_2}^2 \bigg), \ea

where the subscripts on $F$ denote the respective partial
derivatives. Interestingly, the thermodynamic curvature
corresponds to the nature of the correlation present in the
statistical system. For the two component systems \cite{bnt}, the
scalar curvature can be identified as the correlation area $
R(\mu_1, \mu_2) \sim \xi^2 $, where $ \sqrt{\xi(T_C)} $ defines
the correlation length of the corresponding system, at a given
temperature $ T_C(\mu_1, \mu_2) $. \color{black}
\begin{figure*}
\includegraphics[height=.2\textheight,angle=-90]{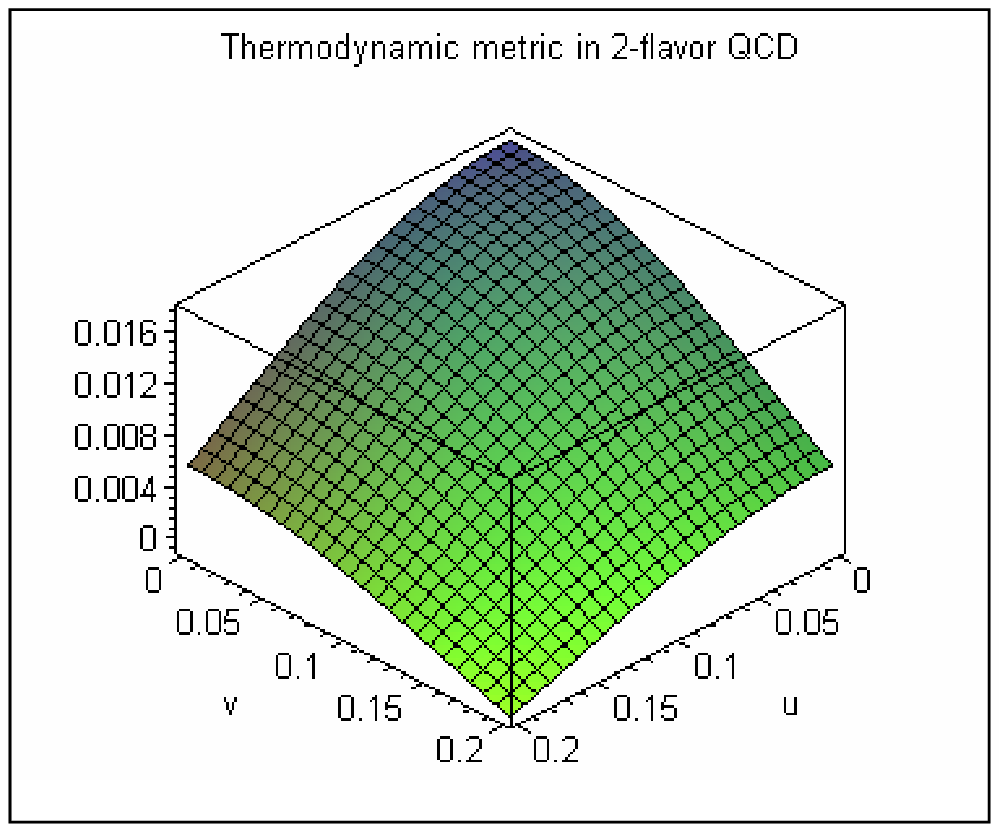}
\includegraphics[height=.2\textheight,angle=-90]{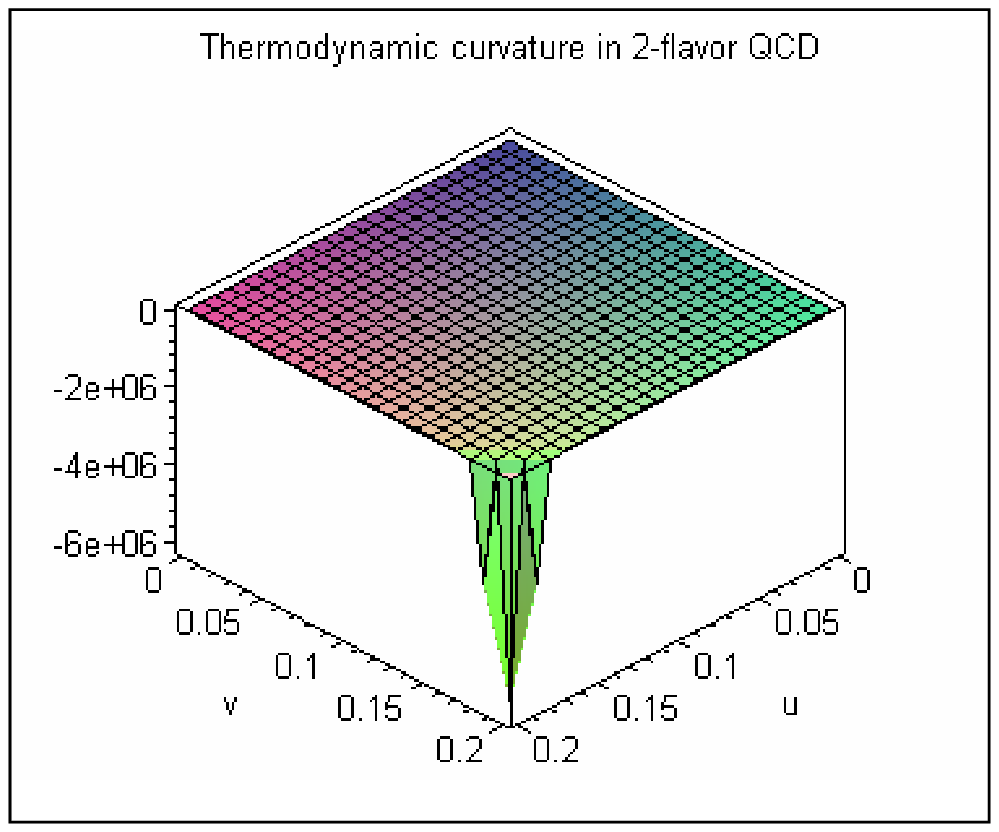}
\caption{The determinant of the thermodynamic metric and scalar
curvature in 2-flavor QCD in the chemical potentials surface. Note
that we measure the chemical potential in GeV.}
\end{figure*}
\section{2-flavor QCD}
Employing Eq.(\ref{eq2}) and Eq.(\ref{eq3}), we find that the
determinant of the thermodynamic metric and the scalar curvature
take the following forms: \ba \label{eq17}
g(\mu_1,\mu_2)&=&\sum_{k,l=0 \vert k+l\le 6}^6 a^{A}_{k,l}
\mu_1^{2 k}\mu_2^{2l}, \nn 
R(\mu_1, \mu_2)&=& -\frac{4}{g^2}\sum_{k,l=0 \vert k+l\le 7}^7
b^{A}_{k,l} \mu_1^{2k}\mu_2^{2l} . \ea Their respective behavior
is shown in the Fig. 1 for a given
range of the interest $(\mu_1, \mu_2)\equiv(u,v)$.

Interestingly, the component of the metric tensor shows symmetry
properties, {\it viz.} symmetry under the exchange of $\mu_1$ and
$\mu_2$, which from the perspective of the quark susceptibility,
turns out to be the parity symmetry. We find that the determinant
of the metric tensor remains non-zero over the domain of chemical
potentials and thus the thermodynamic configuration near $T_C$
remains non-degenerate.

For the 2-flavor QCD, the inclusion of thermal fluctuation leads
to the following positive definite determinant of the metric
tensor and polynomial scalar curvature \ba g(\mu_1,\mu_2)&=& 4
\frac{\sum_{k,l=0\vert k+l\le 7}^7 a^{C}_{l,k}\ \mu_1^{2k}
\mu_2^{2l}}
{(\mu_1^2+\mu_2^2))^{2}(0.20-0.34(\mu_1^2-\mu_2^2))^{7}}, \nn
R(\mu_1, \mu_2)&=& -\frac{4}{g^2} ( 0.20 -0.34 (\mu_1^2
+\mu_2^2)\nn&& \times \sum_{k,l=0 \vert k+l\leq 12}^{12}
a^{C}_{k,l}\ \mu_1^{2 k}\mu_2^{2l}, \ea where the coefficients $
\lbrace a_{i,j} \rbrace $ appearing in the numerator of the
curvature are depicted in the Fig.(2). It is observed that thermal
fluctuations preserve the symmetry properties of the thermodynamic
geometry.
\section{3-flavor QCD}
In this section, we discuss the thermodynamic geometry arising from the
free energy of 3-flavor hot QCD. We obtain the following
expression for the thermodynamic metric and the scalar curvature:
\ba \label{eq29}
 g(\mu_1, \mu_2)&=&
\sum_{k,l=0\vert k+l\le 6}^6 \tilde{a}^{A}_{k,l}\
\mu_1^{2l}\mu_2^{2k}, \nn
R(\mu_1, \mu_2) &=& -\frac{4}{g^2} \sum_{k,l=0\vert k+l\le 7}^7
\tilde{b}^{A}_{k,l}\ \mu_1^{2k}\mu_2^{2l} \ea
\begin{figure*}
\includegraphics[height=.2\textheight,angle=-90]{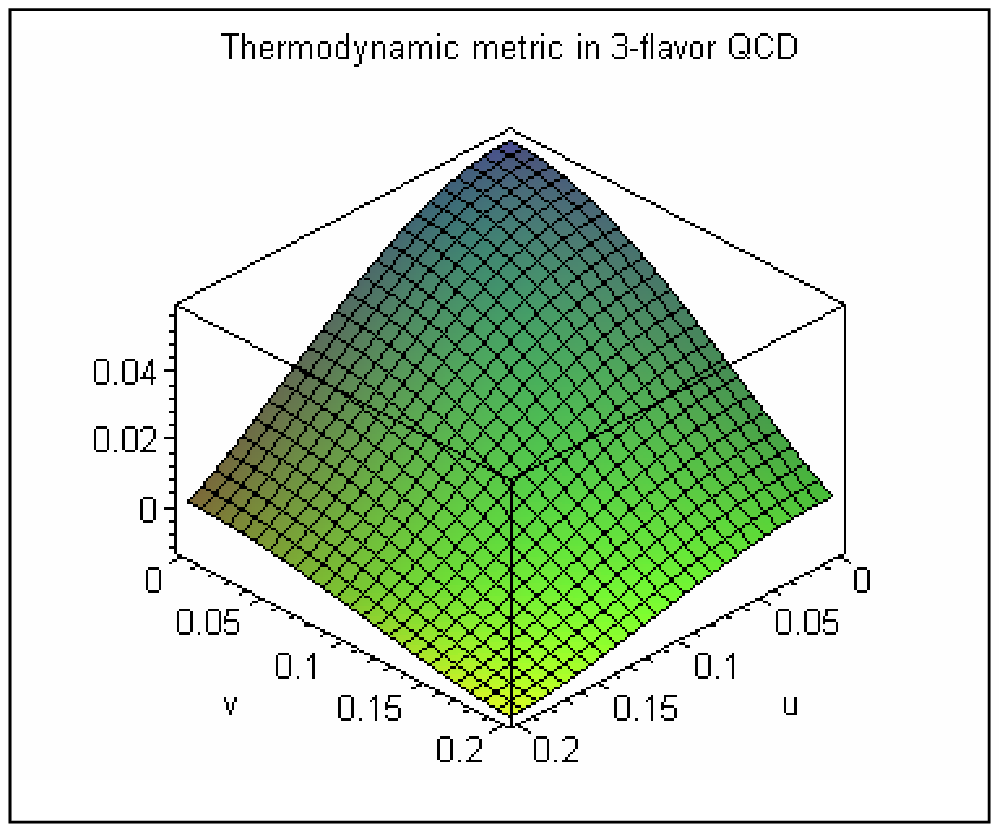}
\includegraphics[height=.2\textheight,angle=-90]{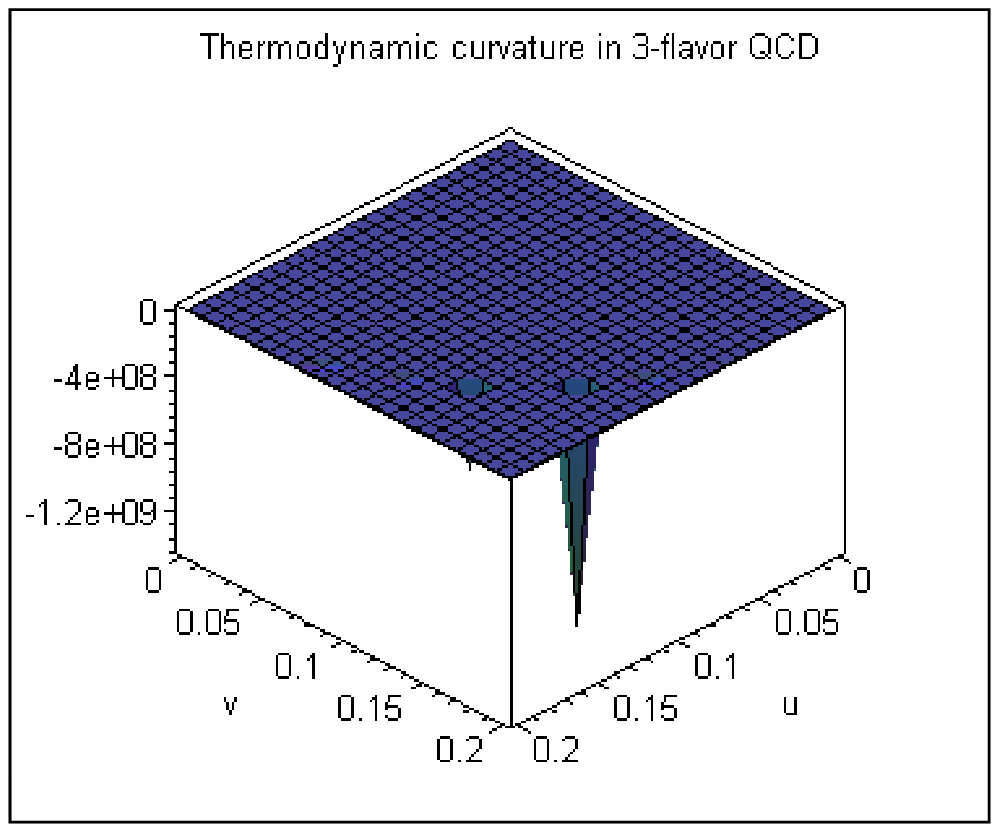}
\caption{The determinant of the thermodynamic metric and scalar
curvature in 3-flavor QCD in the chemical potentials surface. Note
that we measure the chemical potential in GeV.}
\end{figure*}
We find that the components of the metric tensor at zero chemical
potentials take the values $ g_{\mu_1,\mu_1}= 0.26$,
$g_{\mu_2,\mu_2}=0.22 $ and $ g_{\mu_1,\mu_2}= 0 $, which in
effect describe the diagonal and off diagonal quark susceptibility
tensors. At zero chemical potentials, the determinant of the
metric tensor reduces to $g= 0.06$ and the scalar curvature
vanishes identically, which indicates a non-interacting
statistical system.

For the 2-flavor QCD, considering the fact that the free energy is
not symmetric under the exchange of the chemical potentials, we
see for the 3-flavor QCD that $g(\mu_1,\mu_2)\neq g(\mu_2,\mu_1)$.
Subsequently, the underlying diagonal quark susceptibility tensors
differ even for vanishing chemical potentials. The same remains
true for the determinant of the metric tensor and the associated
scalar curvature. In fact, the inclusion of the quartic terms does
not alter our observations. Finally, as for the 2-flavor QCD, we
find that the symmetry of the metric and scalar curvature remains
intact under the inclusion of thermal fluctuations, up to a
refinement \cite{ourpaper}.
\section{Discussion and Conclusion}
In this paper, we have examined an intrinsic geometric notion to
the $2$-flavor and $3$-flavor QCD thermodynamics. The free energy
of the $2$ and $3$-flavor QCD, near $T_c$, with and without the
logarithmic contributions under thermal fluctuations, introduces
the thermodynamic geometric nature of the quark susceptibilities.
Such a consideration is herewith shown to exhibit macroscopic
versus microscopic duality relations via the fermion number
density. These notions are anticipated to be realized as a mixing
between the different flavors/colors, encoded in the corresponding
partition function of the effective gauge field theory. It would
be interesting to understand whether this kind of approach can be
pushed further for QCD, for example in order to predict geometric
properties of chemical correlations, under sizeable higher order
perturbative and non-perturbative contributions.

\end{document}